\documentclass[aps,prd,reprint,groupedaddress,showpacs]{revtex4-1}

\usepackage{dcolumn}
\usepackage{bm}
\usepackage{color}
\usepackage{amsmath}
\usepackage{amssymb}
\usepackage{graphicx}
\usepackage{slashed}
\usepackage[utf8]{inputenc}

\begin{document}


\title{Lorentz violation bounds on Bhabha scattering}

\author{B. Charneski}
\email[]{bruno@fma.if.usp.br }

\author{M. Gomes }
\email[]{mgomes@fma.if.usp.br }

\author{R. V. Maluf}
\email[]{rmaluf@fma.if.usp.br }

\author{A. J. da Silva}
\email[]{ajsilva@fma.if.usp.br }

\affiliation{Instituto de F\'isica, Universidade de S\~ao Paulo \\
 Caixa Postal 66318, 05315-970, S\~ao Paulo, SP, Brazil }

\begin{abstract}
We investigate the effect of Lorentz-violating terms on Bhabha scattering
in two distinct cases correspondent to vectorial and axial nonminimal couplings in QED. In both cases, we find significant modifications with respect
to the usual relativistic result. Our results reveal an anisotropy of the
differential cross section which imply new constraints on the
possible Lorentz violating terms. 
\end{abstract}


\maketitle


\section{\label{sec:intro}Introduction}

Since the Carroll-Field-Jackiw seminal paper \cite{CarrollFieldJackiw}
and after the construction of the extended Standard Model (SME) by
Colladay and Kostelecky  \cite{ColladayKostelecky1,ColladayKostelecky2} (see also \cite{Kostelecky:2000mm} and references therein),
the possibility of Lorentz covariance breakdown in the context of
Quantum Field Theory has been extensively studied. The interest in
this issue appears in different contexts, such as supersymmetric models
\cite{Berger,Carlson:2002zb}, noncommutative geometry \cite{CarrollNoncommutative}, gravity and cosmology \cite{KosteleckyGravity,Collins:2004bp,BLiCosmology,Csaki:2000dm}, high derivative models \cite{Gomes:2009ch,Mariz:2011ed,Gomes:2011pq}, renormalization
\cite{ChenQED,CaroneQED,Charneski:2010mv,Charneski:2010ew} and scattering processes \cite{ColladayScattering,AltschulScattering}
in quantum electrodynamics (QED), condensed matter systems \cite{Belich2005,Belich2006,Charneski},
and so on. Following these theoretical developments, many experimental
tests on Lorentz-violating (LV) corrections have also been carried
out and several constraints on LV parameters were established \cite{MattinglyTest_LV}. One of the most precise experiments, the clock anisotropy, which is a spectroscopic experiment, determines bounds of $10^{-33}$GeV \cite{Brown:2010dt} when LV parameters are introduced as in the SME \cite{ColladayKostelecky1,ColladayKostelecky2}. 
However, for scattering processes, there are few studies about
possible effects of LV on cross sections aimed to determination of upper
bounds on the breaking parameters \cite{ColladayScattering,AltschulScattering,MSchreck}.

In the usual aproach to LV theories, the breaking term is implemented on the kinetic
sector and implies in modifications on the energy-momentum relations,
the free propagators and scattering states as have been stressed in Refs. \cite{ColladayScattering,AltschulScattering}.  An alternative procedure, is to modify just the interactions part via a nonminimal coupling with terms like
$\epsilon_{\mu\nu\alpha\beta}v^{\nu}F^{\alpha\beta}$ and $\epsilon_{\mu\nu\alpha\beta}\gamma_{5}b^{\nu}F^{\alpha\beta}$. In Ref. \cite{Belich2005} this possibility was used used to evaluate the induction of topological phases on fermion systems. Later on, its implication on the spectrum of the hydrogen atom providing the determination
of bounds on the magnitude of the LV coefficients were reported in Ref.
\cite{Belich2006}. However, possible effects on scattering processes
in the framework of QED by these nonminimal couplings have not been
investigated. That is the main objective of this paper, i.e. to obtain a bound to Lorentz violation from a scattering process involving a nonminimal coupling. Bounds obtained from noncolliders experiments \cite{Belich2006} usually depend on the study of the hyperfine structure what is outside of the scope of this work.

Collision experiments in high energy physics provide a suitable environment
where Lorentz symmetry breaking can be tested. Moreover, Bhabha scattering
is one of the most fundamental reactions in QED processes and has
been extensively studied in colliders \cite{VenusCollider,TopazEtAll,RentoSymposium}.
It is particularly important since it is used to determine the luminosity
of the $\mbox{e}^{+}\mbox{e}^{-}$ collisions \cite{MDerrick1986,TArima1997}.
This fact motivated us to evaluate and analise the behavior of the differential cross section
for Bhabha scattering in the presence of nonminimal couplings and to
directly obtain upper bounds on LV coefficients. As we will show,
our calculations can be done similarly to those in standard QED. We
found that the breaking of Lorentz symmetry leads to an unusual dependence
of the cross section on the orientation of the scattering plane in
the center of mass reference frame.

This paper is organized as follows. In Sec. \ref{sec:bhabhaVec},  the differential cross section for Bhabha scattering on the presence of the vectorial nonminimal coupling is calculated. The results obtained are analyzed and a bound to the magnitude of the Lorentz violation is established. In Sec. \ref{sec:bhabhaAxial}, the axial-like nonminimal coupling is considered. In Sec. \ref{sec:concl}, some final remarks are made.

\section{\label{sec:bhabhaVec}Bhabha scattering: vectorial nonminimal coupling }

In this section we calculate the unpolarized differential cross
section for Bhabha scattering $e^{+}e^{-}\rightarrow e^{+}e^{-}$,
in an extended version of QED characterized by a nonminimal covariante derivative \cite{Belich2005,Belich2006}: 
\begin{equation}
D_{\mu}=\partial_{\mu}+ieA_{\mu}+igv^{\nu}F_{\mu\nu}^{\ast},\label{eq:TFNMC}
\end{equation}
where $\ensuremath{F_{\mu\nu}^{\ast}=\frac{1}{2}\varepsilon_{\mu\nu\alpha\beta}F^{\alpha\beta}}$
is the dual electromagnetic tensor with $\ensuremath{\epsilon^{0123}=1}$;
$e$, $g$, $v^{\mu}$ are the electron charge, a coupling constant
and a constant four vector, respectively. With such modification the QED
Lagrangian is 
\begin{eqnarray}
\mathcal{L}&=&-\frac{1}{4}F^{\mu\nu}F_{\mu\nu}+\bar{\psi}(i\gamma^{\mu}\partial_{\mu}-m)\psi -\frac{1}{2\alpha}(\partial_{\mu}A^{\mu})^{2}\nonumber\\
&-&e\bar{\psi}\gamma^{\mu}\psi A_{\mu}-gv^{\nu}\bar{\psi}\gamma^{\mu}\psi\partial^{\alpha}A^{\beta}\epsilon_{\mu\nu\alpha\beta}.\label{eq:QED}
\end{eqnarray}
The additional vertex is gauge invariant, but explicitly violates
Lorentz symmetry, since $v^{\mu}$ defines a privileged direction in the
space-time. Furthermore, it is not perturbatively renormalizable,
since their coupling constant has mass dimension $\left[gv^{\mu}\right]=-1$.

As in standard QED, the Feynman rules can be read directly from Eq.\eqref{eq:QED},
telling us how to write down the tree-level diagrams related in
the process $e^{-}(p_{1})e^{+}(q_{1})\rightarrow e^{-}(p_{2})e^{+}(q_{2})$. In this work we will assume the Feynman gauge $(\alpha=1)$ and the result, to lowest order, for the $S$-matrix element is therefore
\begin{equation}
i\mathcal{M}_{\mbox{total}}=i\mathcal{M}_{0}+i\mathcal{M}_{1}+i\mathcal{M}_{2},\label{eq:Mtotal}
\end{equation}
where $i\mathcal{M}_{0}$ is just the matrix element in conventional
QED:
\begin{eqnarray}
i\mathcal{M}_{0}&=&ie^{2}\left[\frac{\overline{u}(p_{2})\gamma^{\alpha}u(p_{1})\overline{v}(q_{1})\gamma_{\alpha}v(q_{2})}{(p_{1}-p_{2})^{2}}\right.\nonumber\\
&&\left.-\frac{\overline{u}(p_{2})\gamma^{\alpha}v(q_{2})\overline{v}(q_{1})\gamma_{\alpha}u(p_{1})}{(p_{1}+q_{1})^{2}}\right].\label{eq:M0}
\end{eqnarray}
The matrix element $i\mathcal{M}_{1}$ is linear in $(gv^{\mu})$
being formed by an usual vertex and another with the Lorentz-violating
term: 
\begin{eqnarray}
i\mathcal{M}_{1}&=&2egv^{\nu}\epsilon_{\mu\nu\sigma\rho}\left[\frac{(p_{1}-p_{2})^{\sigma}\overline{u}(p_{2})\gamma^{\rho}u(p_{1})\overline{v}(q_{1})\gamma^{\mu}v(q_{2})}{(p_{1}-p_{2})^{2}}\right.\nonumber\\
&&\left.+\frac{(p_{1}+q_{1})^{\sigma}\overline{u}(p_{2})\gamma^{\text{\ensuremath{\rho}}}v(q_{2})\overline{v}(q_{1})\gamma^{\text{\ensuremath{\mu}}}u(p_{1})}{(p_{1}+q_{1})^{2}}\right].\label{eq:M1}
\end{eqnarray}
Finally, $i\mathcal{M}_{2}$ is quadratic in $(gv^{\mu})$ as it results
purely from the Lorentz-violating vertex: 
\begin{eqnarray}
i\mathcal{M}_{2}&=& ig^{2}v^{\gamma}v^{\delta}g^{\kappa\lambda}\epsilon_{\epsilon\delta\tau\lambda}\epsilon_{\omega\gamma\sigma\kappa}\nonumber\\
&&\left[\frac{(p_{1}-p_{2})^{\sigma}(p_{1}-p_{2})^{\tau}\overline{u}(p_{2})\gamma^{\omega}u(p_{1})\overline{v}(q_{1})\gamma^{\epsilon}v(q_{2})}{(p_{1}-p_{2})^{2}}\right.\nonumber \\
 &&-\left.\frac{(p_{1}+q_{1})^{\sigma}(p_{1}+q_{1})^{\tau}\overline{u}(p_{2})\gamma^{\text{\ensuremath{\epsilon}}}v(q_{2})\overline{v}(q_{1})\gamma^{\text{\ensuremath{\omega}}}u(p_{1})}{(p_{1}+q_{1})^{2}}\right].\nonumber\\
 \label{eq:M2}
\end{eqnarray}

To evaluate the cross section, we now compute $\left|i\mathcal{M}_{\mbox{total}}\right|^{2}$,
taking an average over the spin of the incoming particles and summing over the outgoing particles. This can be accomplished using the completeness relations: $\sum u^{s}(p)\overline{u}^{s}(p)=\slashed{p}+m$
and $\sum v^{r}(p)\overline{v}^{r}(p)=\slashed{p}-m$, leading to
traces of Dirac matrices products. We performed these trace calculations, which involves the product of up to eight gamma matrices and the Levi-Civita symbol using the FeynCalc package \cite{Mertig}.
Furthermore, as our main goal is
to consider the behavior of the scattering process in the high energy
limit, we set $p_{1,2}^{2}=q_{1,2}^{2}=m^{2}=0$. This is possible
because the $(gv^{\mu})$ factors are overall on all terms. In this
way, we arrive at the following expression: 
\begin{eqnarray}
\frac{1}{4}\sum_{\mbox{spins}}|\mathcal{M}_{total}|^{2}&=&e^{4}\left(\frac{2(s^{2}+u^{2})}{t^{2}}+\frac{4u^{2}}{st}+\frac{2(t^{2}+u^{2})}{s^{2}}\right)\nonumber\\
&+&\mbox{{\bf A}}(v,p_{1,2},q_{1,2})+\mbox{{\bf B}}(v,p_{1,2},q_{1,2}),
\label{eq:SquareMatrix1}
\end{eqnarray}
with $s$, $t$, and $u$ being the Mandelstam variables. 

The first term in \eqref{eq:SquareMatrix1} consists of the usual squared amplitude
of Bhabha scattering and the second and third terms are the corrections
of second and fourth order in $(gv^{\mu})$, represented by $\mbox{{\bf A}}(v,p_{1,2},q_{1,2})$
and $\mbox{{\bf B}}(v,p_{1,2},q_{1,2})$ respectively. The exact form
of these corrections are lengthy and will not be displayed in detail.
However, we notice that the interference terms of odd order cancel
each other.

\begin{figure*}[p]
\begin{centering}
\includegraphics[scale=0.60]{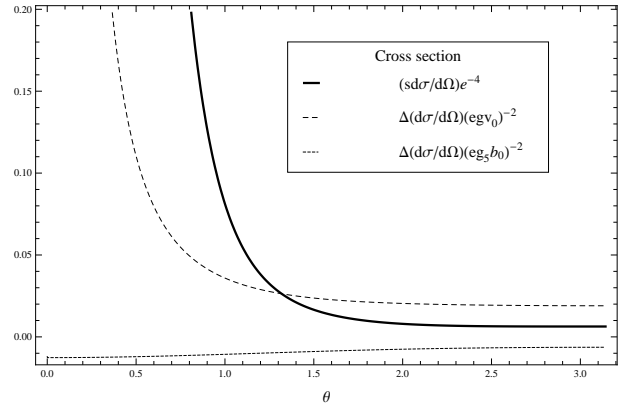} 
\par\end{centering}
\caption{\label{GraficosLorentzViolation}Angular dependence of the differential
cross section for Bhabha scattering to the time-like
Lorentz violation: QED prediction (solid line),
vectorial (dashed line) and axial-like (dotted line) nonminimal couplings.}
\end{figure*}

\begin{figure*}[p]
\begin{centering}
\includegraphics[scale=0.6]{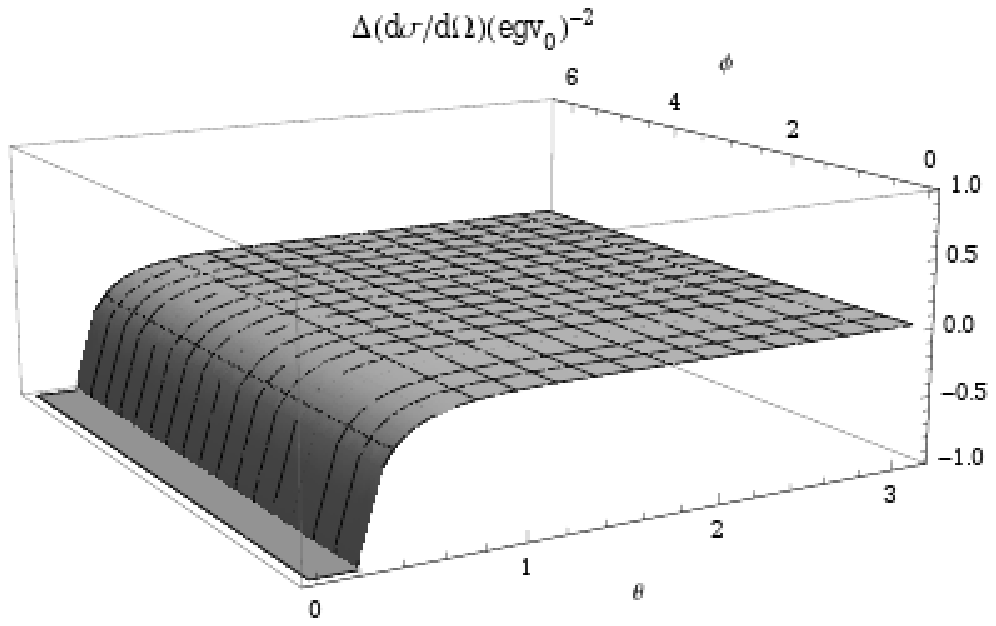}
\includegraphics[scale=0.6]{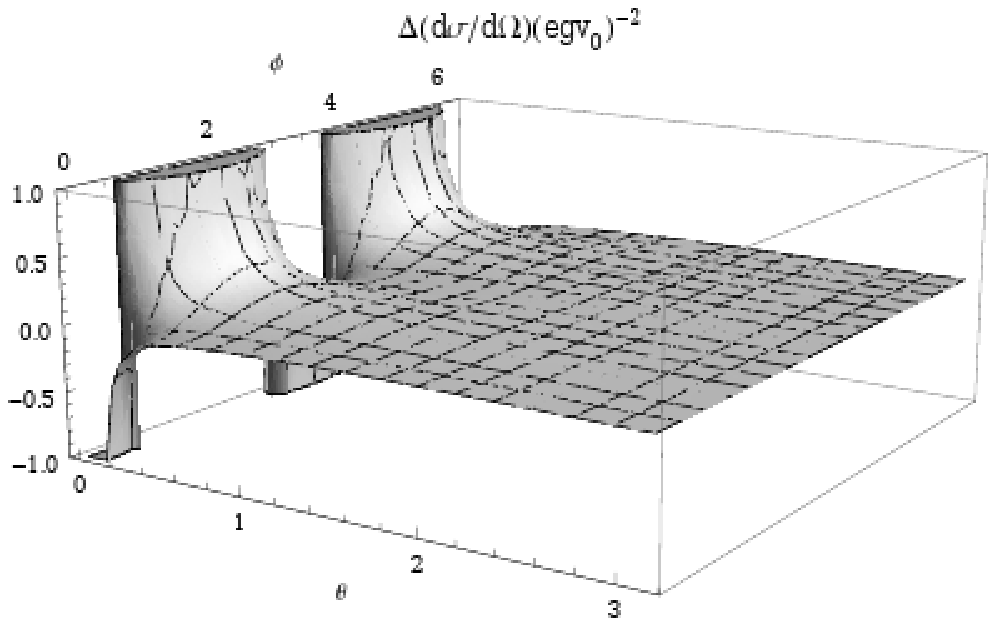}
\par\end{centering}
\caption{\label{GraficosLorentzViolationvSL1}Low order correction to the vectorial cross section (space-like case) for different directions of the background vector: $(\theta_{v}=0,\varphi_{v}=0)$ and $(\theta_{v}=\pi/2,\varphi_{v}=0)$, respectively.}
\end{figure*}

In order to complete the cross section calculation, we must adopt a frame of reference to express the kinematic variables. Bhabha scattering is conventionally analyzed in the center of mass frame, where the 4-momenta take the form
\begin{eqnarray}
p_{1}&=&(E,\ {\bf p}),\ q_{1}=(E,-{\bf p}),\ p_{2}=(E,{\bf \ q}),\nonumber\\
q_{2}&=&(E,-{\bf q}),\ s=(2E)^{2}=E_{cm}^{2},\label{eq:MinkMom}
\end{eqnarray} with ${\bf p}=E{\bf \hat{z}}$, ${\bf q}\cdot\hat{{\bf z}}=E\cos\theta$ and the expression of the differential cross section becomes
\begin{equation}
\frac{d\sigma}{d\Omega}_{\mbox{cm}}=\frac{1}{64\pi^{2}E_{cm}^{2}}.\frac{1}{4}\sum_{\mbox{spins}}|\mathcal{M}_{total}|^{2}.\label{eq:crossSection1}
\end{equation}
We will consider two possibilities according $v^{\mu}$ being time-like or space-like. For the first case where $(v^{\mu}=v_0,0)$ is 
time-like , we can simplify \eqref{eq:SquareMatrix1}
and make use of \eqref{eq:crossSection1}, to obtain 
\begin{widetext}
\begin{eqnarray}
\frac{d\sigma}{d\Omega}_{cm} & = & \frac{e^{4}(\cos2\theta+7)^{2}}{256\pi^{2}E_{cm}^{2}(\cos\theta-1)^{2}}+\frac{v_{0}^{2}g^{2}e^{2}\sin^{2}\frac{\theta}{2}(-65\cos\theta+6\cos2\theta+\cos3\theta+122)}{256\pi^{2}(\cos\theta-1)^{2}}\nonumber \\
 & + & \frac{v_{0}^{4}g^{4}E_{cm}^{2}\sin^{4}\frac{\theta}{2}(-4\cos\theta+\cos2\theta+11)}{128\pi^{2}(\cos\theta-1)^{2}},\label{eq:crossSectionV0}
\end{eqnarray}
\end{widetext}
where the first term is the usual QED differential cross section at lowest order and the second and third terms contain
the contributions of the LV background. This result shows that the differential cross section remains symmetrical
with respect to the colliding beams and its assymptotic angular dependence is qualitatively the same as the usual, as can be seen in Fig.\ref{GraficosLorentzViolation}.
For the second case of interest, we consider $(v^{\mu}=0,\mbox{\textbf{v}})$ space-like  and assuming an arbitrary direction. In this way, we can write the scalar product of vectors as follows: 
\begin{eqnarray}
{\bf p}\cdot{\bf v}&=&E\mbox{v}\cos(\theta_{v}), \nonumber\\
{\bf q}\cdot{\bf v}&=&E\mbox{v}(\sin\theta\sin\theta_{v}\cos(\varphi-\varphi_{v})+\cos\theta\cos\theta_{v})\nonumber\\
&\equiv &E\mbox{v}\cos(\Psi),\label{eq:vecv}
\end{eqnarray}
Thus, after some algebraic simplifications, we get 
\begin{widetext}
\begin{eqnarray}
\frac{d\sigma}{d\Omega}_{cm} & = & \frac{e^{4}(\cos2\theta+7)^{2}}{256\pi^{2}E_{cm}^{2}(\cos\theta-1)^{2}}\nonumber \\
 & + & \frac{e^{2}g^{2}\text{v}^{2}}{256\pi^{2}(\cos\theta-1)^{2}}\left[2(32\cos\theta+\cos2\theta-49)\cos^{2}\frac{\theta}{2}\left(\cos^{2}\theta_{v}+\cos^{2}\Psi\right)\right.\nonumber \\
 & + & 2(61\cos\theta-22\cos2\theta+3\cos3\theta-10)\cos\Psi\cos\theta_{v}\nonumber \\
 & + & \left.\sin^{2}\theta(-64\cos\theta+3\cos2\theta+85)\right]\nonumber \\
 & + & \frac{g^{4}\text{v}^{4}E_{cm}^{2}}{1024\pi^{2}(\cos\theta-1)^{2}}\left[-4(8\cos\theta+\cos2\theta+7)\cos^{3}\Psi\cos\theta_{v}+\right.\nonumber \\
 & + & 2\cos^{2}\Psi\left((12\cos\theta+7\cos2\theta+29)\cos^{2}\theta_{v}-2\sin^{2}\theta(3\cos\theta+7)\right)\nonumber \\
 & - & 4\cos\Psi\cos\theta_{v}\left((8\cos\theta+\cos2\theta+7)\cos^{2}\theta_{v}-4\sin^{2}\theta(3\cos\theta+2)\right)\nonumber \\
 & + & (4\cos\theta+\cos2\theta+11)\cos^{4}\theta_{v}-4\sin^{2}\theta(3\cos\theta+7)\cos^{2}\theta_{v}\nonumber \\
 & + & \left.(4\cos\theta+\cos2\theta+11)\cos^{4}\Psi+8\sin^{4}\frac{\theta}{2}(24\cos\theta+7\cos2\theta+25)\right].\label{eq:crossSectionVecV}
\end{eqnarray}
\end{widetext}
In the above result, we note the dependence of the cross section with
respect to the azimuthal angle $\varphi$. For the fixed background
${\bf v}$ perpendicular to the beam collision $(\theta_v=\pi/2)$, this effect is maximal
and it is characterized by a set of periodic sharp peaks, as illustrated
in Fig.\ref{GraficosLorentzViolationvSL1}. For the Compton scattering with the LV term in the kinetic sector a similar result was reported \cite{MSchreck}.

To conclude this section, we will determine upper bounds for the products
of the parameters $(gv^{\mu})$ in the cases evaluated above. Our
choice to study Bhabha scattering was motivated, in addition to the
questions outlined in the introduction, by practical reasons, i.e,
the experimental data on precision tests for this kind of scattering
in QED are readily available in Ref. \cite{MDerrick1986}. In the
experiment reported in that paper, the measurements of the differential
cross sections for $e^{+}e^{-}\rightarrow e^{+}e^{-}$ and $e^{+}e^{-}\rightarrow\gamma\gamma$
scatterings were evaluated at a center-of-mass energy of 29 GeV and
in the polar-angular region $\left|\cos\theta\right|<0.55$. For Bhabha
scattering, small deviations on the magnitude of the QED tree results may be expressed
in the form: 
\begin{equation}
\Bigg{|}\left.\left(\frac{d\sigma}{d\Omega}\right)\right/\left(\frac{d\sigma}{d\Omega}\right)_{QED}-1\Bigg{|}\approx\left(\frac{3s}{\Lambda^{2}}\right),\label{eq:LambdaCorrections}
\end{equation}
where $s=E_{cm}^{2}$ and $\Lambda$ is a small parameter representing possible experimental departures from the theoretical predictions (see Table XIV of Ref. \cite{MDerrick1986}).

Considering the leading corrections for small ($gv^{\mu}$) in \eqref{eq:crossSectionV0}
and \eqref{eq:crossSectionVecV}, we can show that the magnitude of
these corrections are of order $g^{2}v^{2}s/e^{2}$, and therefore
when compared with \eqref{eq:LambdaCorrections} may not be larger
than $3s/\Lambda^{2}$. Thus, we obtain the upper bound 
\begin{equation}\label{boundprimo}
(gv^{\mu})\leq10^{-12}(\mbox{eV})^{-1},
\end{equation}
for $\Lambda=200$ GeV. 

In the above calculations we provided a way to obtain bounds to LV from the analyses of the Bhabha scattering experiment using only QED interactions. The inclusion of QCD effects would improve the value of $\Lambda$ (consequently the bound) and should allow a better comparison with the results encountered for atomic clocks or torsion balances.

\section{\label{sec:bhabhaAxial}Bhabha scattering: axial-like nonminimal coupling}

We turn our attention now to the nonminimal coupling of chiral character,  defined as
\begin{equation}
D_{\mu}=\partial_{\mu}+ieA_{\mu}+ig_{5}\gamma^{5}b^{\nu}F_{\mu\nu}^{\ast},\label{eq:TLNMC}
\end{equation}
which was also examined in Refs. \cite{Belich2005,Belich2006}.

The calculation of the unpolarized cross section may be worked out
similarly as in the previous section. Note that the expression for $i\mathcal{M}_{2}$ differ from \eqref{eq:M2} just for the insertion of the $\gamma_5$ matrix in each matrix element, whereas for $i\mathcal{M}_{1}$ we have the mixture of the vertices:
\begin{eqnarray}
i\mathcal{M}_{1}&=&\frac{eg_{5}b^{\nu}(p_{1}-p_{2})^{\theta}\epsilon_{\mu\nu\theta\rho}}{(p_{1}-p_{2})^{2}}\left[u(p_{2})\gamma^{\rho}u(p_{1})\overline{v}(q_{1})\gamma^{\mu}\gamma^{5}v(q_{2})\right.\nonumber\\
&&\left.-\overline{u}(p_{2})\gamma^{\text{\ensuremath{\mu}}}\gamma^{5}u(p_{1})\overline{v}(q_{1})\gamma^{\text{\ensuremath{\rho}}}v(q_{2})\right]\nonumber \\
 &&+\frac{eg_{5}b^{\nu}(p_{1}+q_{1})^{\theta}\epsilon_{\mu\nu\theta\rho}}{(p_{1}+q_{1})^{2}}\left[\overline{u}(p_{2})\gamma^{\text{\ensuremath{\rho}}}v(q_{2})\overline{v}(q_{1})\gamma^{\text{\ensuremath{\mu}}}\gamma^{5}u(p_{1})\right.\nonumber\\
 &&\left.-\overline{u}(p_{2})\gamma^{\text{\ensuremath{\mu}}}\gamma^{5}v(q_{2})\overline{v}(q_{1})\gamma^{\text{\ensuremath{\rho}}}u(p_{1})\right].
\end{eqnarray}

In the high energy limit and the center of mass frame the differential
cross section for the case $b^{\mu}=(b_{0},0)$ is given by
\begin{eqnarray}
\frac{d\sigma}{d\Omega}_{cm}&=&\frac{e^{4}(\cos2\theta+7)^{2}}{256\pi^{2}E_{cm}^{2}(\cos\theta-1)^{2}}\nonumber\\
&-&\frac{b_{0}^{2}g_{5}^{2}e^{2}\sin^{4}\frac{\theta}{2}(8\cos\theta+\cos2\theta+23)}{64\pi^{2}(\cos\theta-1)^{2}}\nonumber\\
&+&\frac{b_{0}^{4}g_{5}^{4}E_{cm}^{2}\sin^{4}\frac{\theta}{2}(-4\cos\theta+\cos2\theta+11)}{128\pi^{2}(\cos\theta-1)^{2}}.
\end{eqnarray}
Similarly to the time-like case evaluated in the previous section,
the above result contains only even terms in $(g_{5}b_{0})$.
However, the asymptotic behavior is quite different and the leading-order
contribution is finite in the limit $\theta\rightarrow0$, as shown
by the dotted line in Fig.\ref{GraficosLorentzViolation}.

\begin{figure*}[p]
\begin{centering}
\includegraphics[scale=0.6]{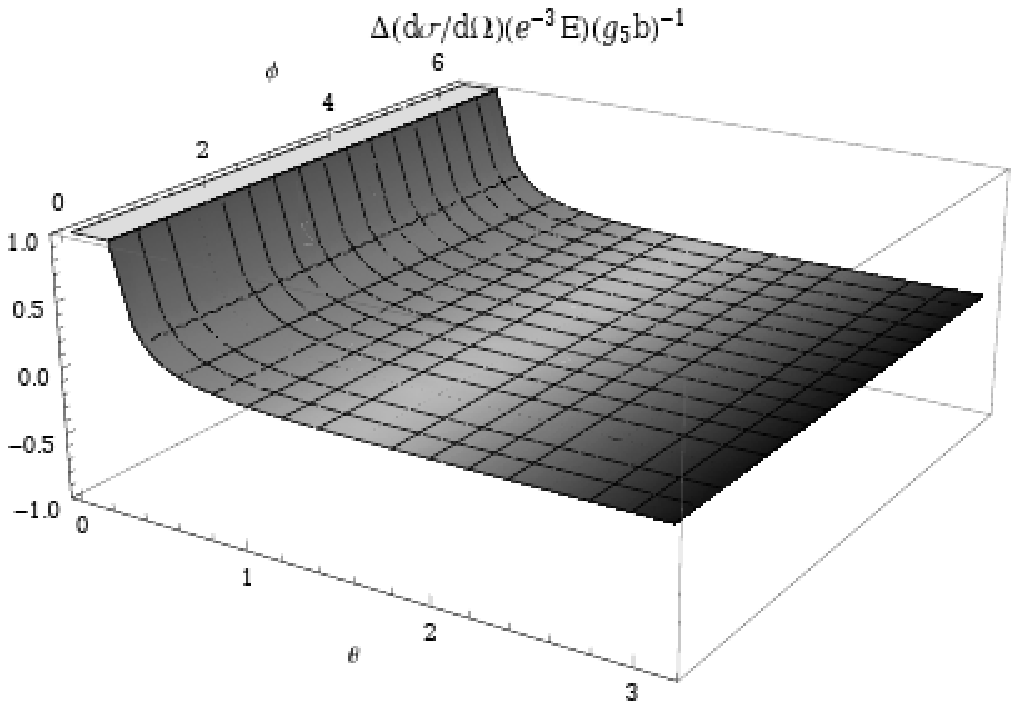}
\includegraphics[scale=0.6]{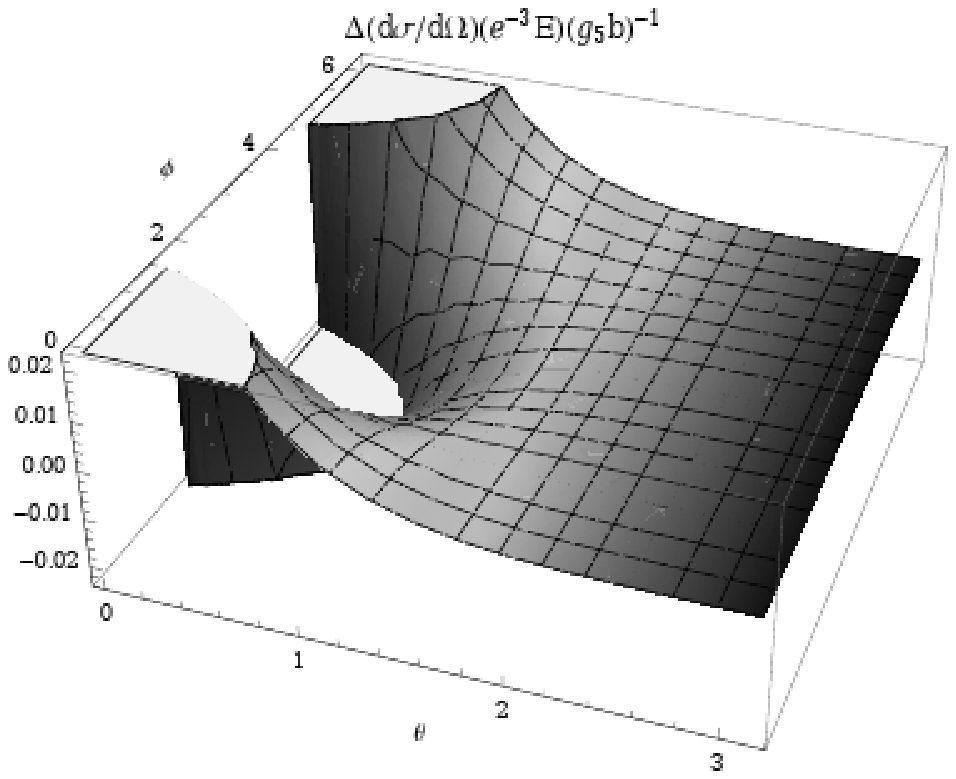}
\includegraphics[scale=0.6]{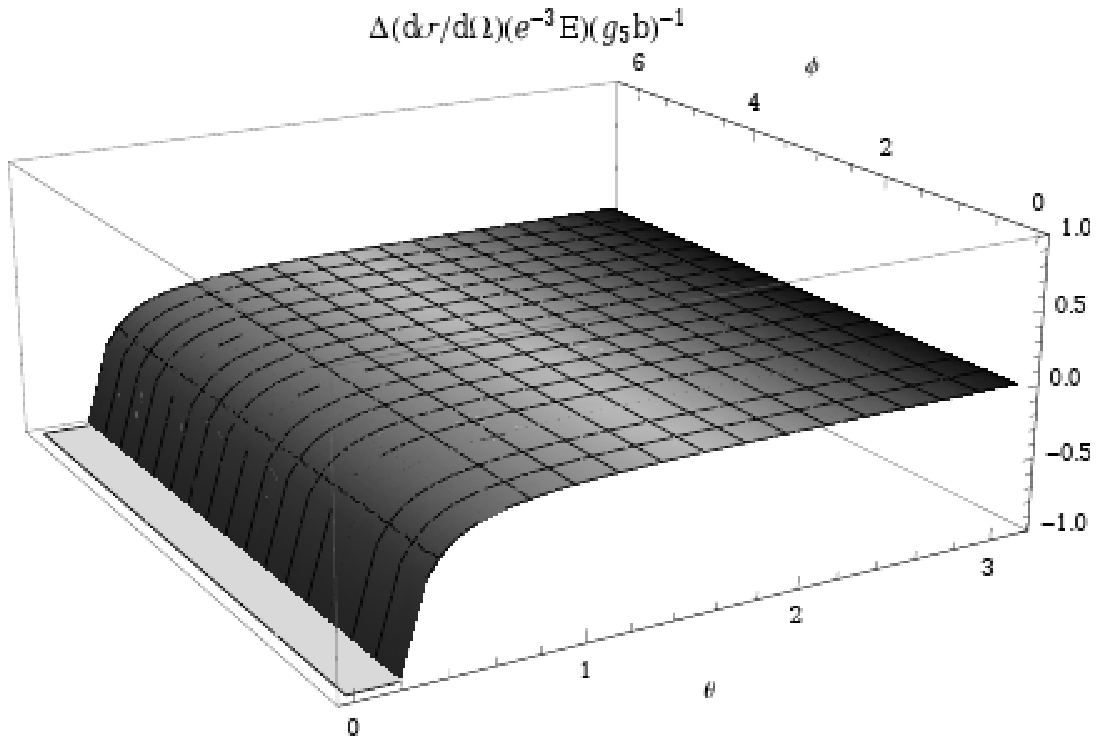}
\end{centering}
\caption{\label{GraficosLorentzViolationvSL5}Low order correction to the axial-like cross section (space-like case) for different directions of the background vector: $(\theta_{5}=0,\varphi_{5}=0)$, $(\theta_{5}=\pi/2,\varphi_{5}=0)$ and $(\theta_{5}=\pi,\varphi_{5}=0)$, respectively.}

\end{figure*}

For the case $b^{\mu}=(0,{\bf b})$, the differential cross section
becomes
\begin{widetext}
\begin{eqnarray}
\frac{d\sigma}{d\Omega}_{cm} & = & \frac{e^{4}(\cos2\theta+7)^{2}}{256\pi^{2}E_{cm}^{2}(\cos\theta-1)^{2}}\nonumber \\
 & + & \frac{e^{3}g_{5}\text{b}\sin^{2}\theta(\cos\theta+1)(\cos\theta_{5}+\cos\Psi_{5})}{16\pi^{2}E_{cm}(\cos\theta-1)^{2}}\nonumber \\
 & + & \frac{e^{2}g_{5}^{2}\text{b}^{2}}{512\pi^{2}(\cos\theta-1)^{2}}\left[8(7\cos\theta-6\cos2\theta+\cos3\theta+14)\cos\Psi_{5}\cos\theta_{5}\right.\nonumber \\
 & + & (-\cos\theta+26\cos2\theta+\cos3\theta-90)(\cos^{2}\theta_{5}+\cos^{2}\Psi_{5})\nonumber \\
 & + & \left.6\sin^{2}\theta(-24\cos\theta+\cos2\theta+31)\right]\nonumber \\
 & + & \frac{eg_{5}^{3}\text{b}^{3}E_{cm}\sin^{2}\theta}{32\pi^{2}(\cos\theta-1)^{2}}\left[3(\cos\theta-1)+(\cos\Psi_{5}-\cos\theta_{5}){}^{2}\right](\cos\theta_{5}+\cos\Psi_{5})\nonumber \\
 & + & \frac{g_{5}^{4}\text{b}^{4}E_{cm}^{2}}{1024\pi^{2}(\cos\theta-1)^{2}}\left[-4(8\cos\theta+\cos2\theta+7)\cos^{3}\Psi_{5}\cos\theta_{5}\right.\nonumber \\
 & + & 2(12\cos\theta+7\cos2\theta+29)\cos^{2}\Psi_{5}\cos^{2}\theta_{5}\nonumber \\
 & + & 4\cos\Psi_{5}\cos\theta_{5}\left(4\sin^{2}\theta(3\cos\theta+2)-(8\cos\theta+\cos2\theta+7)\cos^{2}\theta_{5}\right)\nonumber \\
 & - & 2\sin^{2}\theta(3\cos\theta+7)\left(\cos2\theta_{5}+\cos2\Psi_{5}+2\right)\nonumber \\
 & + & (4\cos\theta+\cos2\theta+11)\cos^{4}\theta_{5}+(4\cos\theta+\cos2\theta+11)\cos^{4}\Psi_{5}\nonumber \\
 & + & \left.8\sin^{4}\frac{\theta}{2}(24\cos\theta+7\cos2\theta+25)\right],
\end{eqnarray}
\end{widetext}
where the presence of odd-order corrections in $(g_{5}\mbox{b})$,
absent in the previous vectorial cases is to be noticed. Furthermore, the effect
of anisotropy in the cross-section is highlighted,
as indicated in Fig. \ref{GraficosLorentzViolationvSL5}.

Now, an analyses similar to the previous section allows to set up an
upper bound to the breaking parameter $(g_{5}b^{\mu})$. Taking into account the
magnitude of the leading-order corrections for the time-like and space-like
cases given respectively by $4g_{5}^{2}b_{0}^{2}s/e^{2}$
and $16g_{5}\mbox{b}\sqrt{s}/e$ and assuming that $s=29$ GeV and $\Lambda_{\pm}=200$
GeV, we find 
\begin{equation}
g_{5}b_{0}\leq10^{-12}(\mbox{eV})^{-1}\ \ \ \ \mbox{and}\ \ \ \ g_{5}\mbox{b}\leq10^{-14}\mbox{(eV)}^{-1}.
\end{equation}

\section{\label{sec:concl}Conclusion}

In this paper, the implications of Lorentz symmetry breaking on Bhabha
scattering have been studied. The LV background terms were introduced
by nonminimal couplings between the fermion and gauge fields. We calculated the differential cross sections for the vector
and axial couplings and determined upper bounds on the magnitude
of the corresponding LV coefficients, by making use of accurate experimental
data, available in the literature. In particular, when we consider
the vector backgrounds, $v^{\mu}, b^{\mu}$, 
as being purely spatial, the cross section acquires a nontrivial dependence
on the direction of these vectors.Finally, we hope that these results may be usefull as a guide in the investigation of the Lorentz violation phenomena in high energy scattering processes.

\begin{acknowledgments}
This work was partially supported by Conselho Nacional de Desenvolvimento
Cient\'{\i}fico e Tecnol\'ogico (CNPq), and Funda\c{c}\~ao de Amparo \`a  Pesquisa
do Estado de S\~ao Paulo (FAPESP).
\end{acknowledgments}



\end{document}